\documentclass[10pt,twocolumn]{article}
\usepackage{graphicx}
\usepackage{amsfonts}
\usepackage{amsmath}
\usepackage{amssymb}
\usepackage{mathrsfs} 
\usepackage{lipsum}
\usepackage{color}
\usepackage{siunitx}
\usepackage{authblk}
\usepackage[margin=2cm]{geometry}
\usepackage{mathtools, cuted}

\usepackage[T1]{fontenc}
\usepackage{graphicx}
\usepackage{hyperref}

\usepackage{bm}

\begin{document}

\vspace{10pt}

\title{A new setup for giant soap films characterization}

\author{Sandrine Mariot \footnote{Corresponding author}, Marina Pasquet, Vincent Klein, Fr\'ed\'eric Restagno, Emmanuelle Rio\\
\small{Universit\'e Paris-Saclay, CNRS, Laboratoire de physique des solides, 91405 Orsay}}
\date{\today}
\twocolumn[
    \begin{@twocolumnfalse}
        \maketitle
           \begin{abstract}
Artists, using an empirical knowledge, manage to generate and play with giant soap films and bubbles. Until now, scientific studies of soap films generated at a controlled velocity and without any feeding from the top, studied films of a few square centimeters. The present work aims to present a new setup to generate and characterize giant soap films (2~m $\times$ 0.7~m). Our setup is enclosed in a humidity-controlled box of 2.2~m high, 1~m long and 0.75~m large. Soap films are entrained by a fishing line withdrawn out of a bubbling solution at various velocities. We measure the maximum height of the generated soap films, as well as their lifetime, thanks to an automatic detection. This is allowed by light-sensitive resistors collecting the light reflected on the soap films and ensures robust statistical measurements. In the meantime, thickness measurements are performed with a UV-VIS-spectrometer, allowing us to map the soap films thickness over time.
           \end{abstract}

    \end{@twocolumnfalse}]

\section{Introduction}

Artists combine formulation and practice to generate giant soap films and play with their shapes and bright colors \cite{URL_PierreYves}. They manage thereby to generate huge bubbles, up to several meters high: the world record for the tallest free-standing soap bubble is 7.004 meters high \cite{URL_RecordBulleGeante}.
Such big and fragile objects are not only fascinating, they also raise specific scientific questions important for industrial aspects, as well as for the global comprehension of the phenomenon: what is the role of physical chemistry on the stability of the films? What are the role of inertia and gravity on the foam film generation and stability?

\begin{figure*}[htbp]
    \centering
    \def\svgwidth{1\textwidth}
    \includegraphics{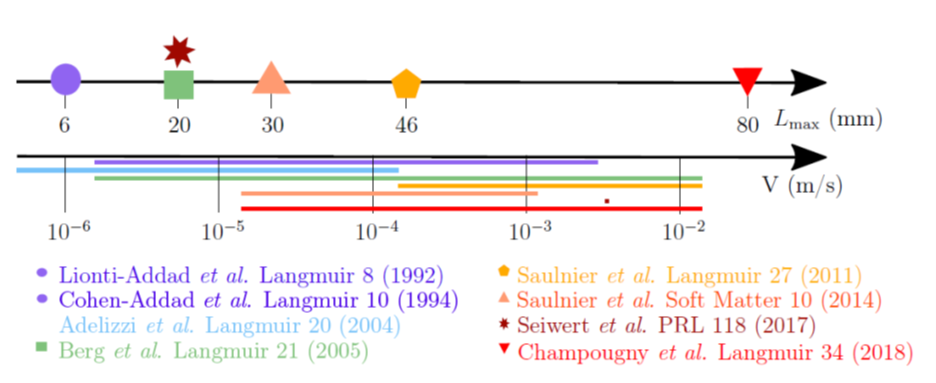}
    \caption{Schematic diagram of the range of maximum lengths and velocities studied in the literature for soap films generated by pulling a frame out of a liquid bath. }
    \label{fig:Velocity_Range}
\end{figure*}

Up to now, most experimental studies have been focused on soap films generation at small scales, up to less than 5~cm high, as we can see in Fig. \ref{fig:Velocity_Range} \cite{adelizzi_interfacial_2004,cohen-addad_stabilization_1994,Saulnier2011,Saulnier2014,berg_experimental_nodate,adami_capillary-driven_2014,seiwert_velocity_2017,adami_capillary-driven_2014}. These soap films were generated at low velocities, less than a few centimetres per second. Lionti-Addad \textit{et al.} \cite{lionti-addad_stabilization_1992}, Cohen-Adad \textit{et al.} \cite{cohen-addad_stabilization_1994},  Adelizzi \textit{et al.} \cite{adelizzi_interfacial_2004} and Berg \textit{et al.} \cite{berg_experimental_nodate} have studied the withdrawal of entrained films with different surfactants and hydro-soluble polymers. They have measured film thicknesses $h$ by light reflectivity, in order to investigate the area of validity of Frankel's law. For a solution of viscosity $\eta$ and surface tension $\gamma$, the entrained film thickness predicted by the Frankel's law \cite{Mysels1959}  is: 

\begin{equation} 
    h_F =1.89\kappa^{-1} \text{Ca}^{2/3}
    \label{eq.1}
\end{equation}
\noindent where  $\kappa^{-1}=(\gamma / \rho g)^{1/2}$ is the capillary length and  $\text{Ca} = \eta V/ \gamma$ the capillary number. Van Nierop \textit{et al.} \cite{VanNierop2008} have compiled experimental data from the literature on the thickness of soap films as a function of their entrained velocity. They have shown that the Frankel's law in $\text{Ca}^{2/3}$ corresponds well to the experimental data, nevertheless deviations from this law appear at ``large'' capillary number $\text{Ca} \gtrsim 10^{-3}$. In this article we will show that we can reach much larger values of Ca.

More recently, Saulnier \textit{et al.} \cite{Saulnier2011,Saulnier2014} have studied entrained small films stabilized either by C$_{12}$E$_6$ (hexaethylene glycol monododecyl ether) or by SDS (sodium dodecylsulfate). They have studied the rupture of the films created by a frame pulled out of a liquid bath \cite{Saulnier2014}. They obtained the same lifetime for the different surfactants, and conclude that the films maximum length does not depend on the physico-chemistry. On the other hand, the artists who make giant bubbles pay attention to the composition of the solutions they use: with this new setup, we will be able to work under controlled conditions and see whether or not the solutions play a role in the stability of the giant films.

The aim of this article is to present a new setup to generate giant soap films at controlled velocities, comparable to those apply by the artists when they create bubbles by hand. We thus need to increase by a factor of 300 the pulling velocity compared to previous experiments  \cite{adelizzi_interfacial_2004,cohen-addad_stabilization_1994,Saulnier2011,Saulnier2014}. Until now, no study has been carried out during films generation at such velocities (Fig. \ref{fig:Velocity_Range}). We also need to increase the size of the soap films by more than a factor of 40 compared to the maximum size reported in the literature. Note that the question that we want to arise here is the stability of giant soap films during and after their generation, which is a problem different than the stability of films fed from the top. The later are called  ``curtains'' in the literature, and many studies have been carried out to characterize them \cite{adami_surface_2015, Kim2017, Sane2018}, to use them as a support to study two-dimensional problems such as turbulence \cite{Rutgers2001, Kellay2002, kellay_hydrodynamics_2017} or to investigate three-dimensional hydrodynamic problems (impact of liquid jets on the films \cite{Basu2017}, formation of bubbles \cite{Salkin2016}, ...). The largest curtain made in a laboratory had a height of 20~m by a width of 4~m \cite{Rutgers2001}, much longer than any laboratory non fed foam film.  

We present a new setup allowing to create 2 m height soap films with a width of 0.7 m. We reach high entrained velocity, from 0.02 m/s up to 2.5 m/s by combining efficient motors, meaning that the capillary number Ca goes up to 10$^{-1}$ for typical aqueous solutions. This will be fully described in the section \ref{section_SETUP}. In section \ref{Section_Controlled-environnement}, we will see that we can work in a controlled environment where the humidity is regulated and measured. Indeed, knowing the ambient humidity during the generation is important and few experiments have been done in this way \cite{li_effect_2010,li_effect_2012,Champougny2018_evaporation,miguet_stability_2020}, whereas Champougny \textit{et al.} \cite{Champougny2018_evaporation} reported in a recent work that the evaporation has a huge impact on the film rupture. In section \ref{Section_Measurements}, we will describe all the measurements allowed by this setup. In particular, we will show that the automatization allows us to  have a robust statistical analysis and accuracy, which is necessary as shown by Tobin \textit{et al.} \cite{tobin_public_2011}.

\section{Setup} \label{section_SETUP}

\subsection{Mechanical parts} \label{section_Mechanical-parts}

\begin{figure}[htbp]
    \centering
    \includegraphics[width=\columnwidth]{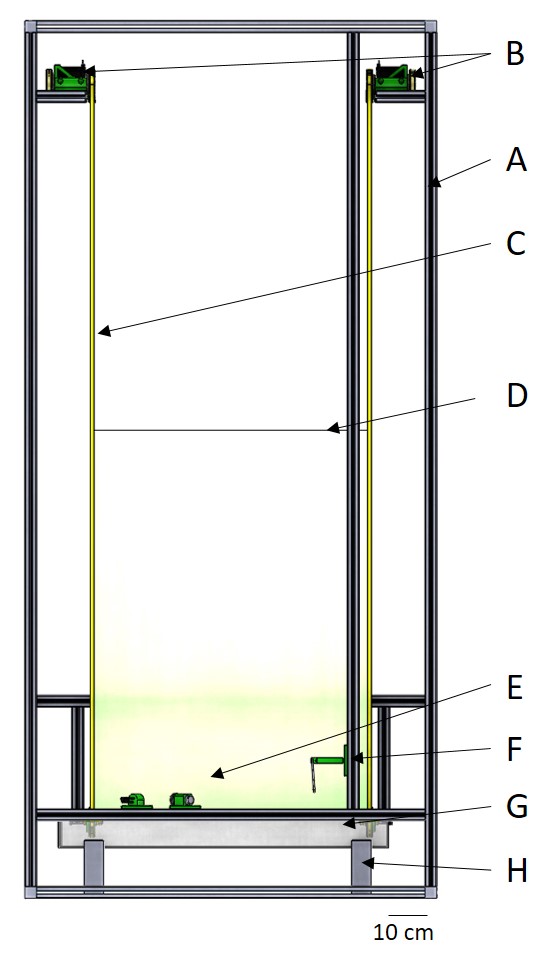}
    \caption{Overall device shown by a Solid Works drawing. The frame (A) is composed of Dural profile and can be closed to control the humidity. At the top, coupled motors (B) are combined to pulleys which entrain toothed drive belts (C) that cause a linear motion to the fishing wire (D). At the bottom, a parallelepipedic glassy tank (G), lying on stands (H), contains the soapy solution. A white light spot coupled to 3 light-sensitive resistors (E) is used to detect the film presence and a UV-VIS spectrometer (F) to study the thickness. }
    \label{fig:schemamanip}
\end{figure}

\begin{figure}[htbp]
    \centering
    \includegraphics[width=\columnwidth]{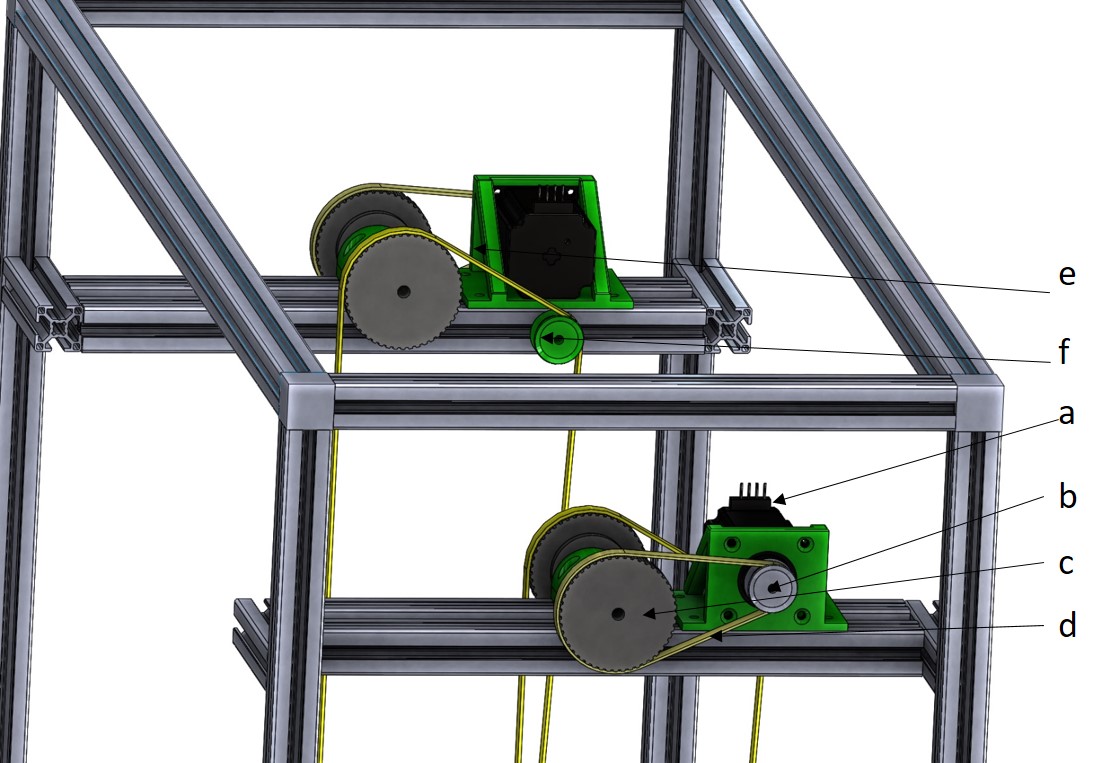}
    \caption{The stepping motor (a) entrains a first pulley (b), diameter 25.06 mm, that is coupled to a second pulley (c), diameter 70 mm,  thanks to a band (d). It allows to multiply the speed.
The bands (e) strain leading  the soap film is set by a round 3D printed piece (f).}
    \label{fig:schemamanip moteur}
\end{figure}

The overall setup that we have designed, shown in Fig. \ref{fig:schemamanip}, is composed of 30 $\times$ 30~mm$^2$ aluminium square profiles (Bosh Rexroth purchased from Radiospare) assembled together as a parallelepiped with overall dimensions of 2.2~m $\times$ 1~m $\times$ 0.75~m. At the top, 2 coupled motors (Fig. \ref{fig:schemamanip moteur}), combined with pulleys, entrain two toothed drive belts, 1 cm $\times$ 4.03~m (purchased from CourroiesConcept) at a constant and controlled velocity. This allows to transform the rotational movement of the motors into a vertical translation for the belts. As the initiator of the soap film, a fishing wire, with a diameter of 0.74~mm (made in fluorocarbon) is fixed perpendicularly to the belts. The soap films are created when the fishing line is withdrawn out of the bubbling solution.

\begin{figure}[htbp]
    \centering
    \includegraphics[width=\columnwidth]{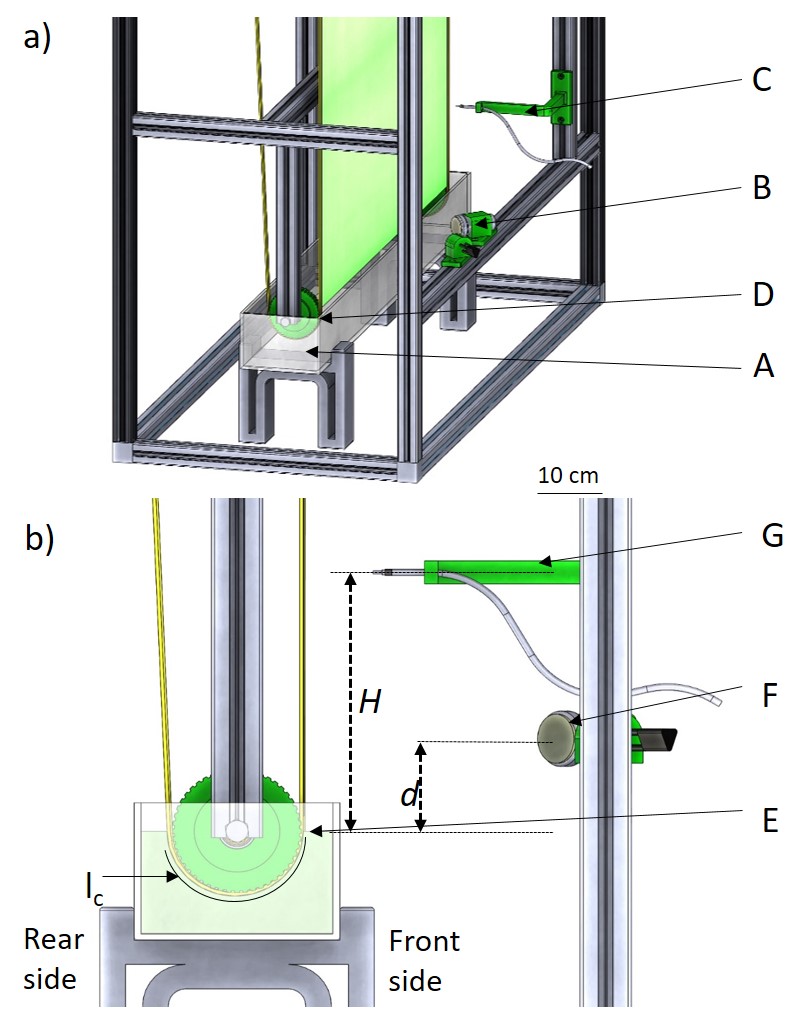}
    \caption{(a) 3-D view of the bottom of the frame. A glassy tank (A) contains the soapy solution. A white light spot coupled to 3 photodiodes (B) is used to detect the film presence and a UV-VIS spectrometer (C) to study the thickness. The fishing line is entrained by  the movement of the pulleys (D). 
    (b)Side drawing of the trough. Half of the pulley is immersed in the soapy solution (A). (B) shows the position where the film is generated. \textit{d} represents the fixed distance between the lighting (B) and the beginning of the film at the surface. \textit{H} represents the variable distance of the spectrometer (D) along the height of the film. $l_{c}$, the circular length, is fixed into the program.}
    \label{fig:schema Side Trough}
\end{figure}

The combined stepping motors (SM56 3 18 J4.6 and associated variators, purchased from Rosier Mecatronique, holding torque ranging from 0.3 to 38~Nm) shown in Fig. \ref{fig:schemamanip moteur} (a) are chosen so that we can control the acceleration and deceleration of the rotational axis. The acceleration time of the motors is 100~ms. Attached to the rotational axe of the stepping motor, a pulley with an external diameter of 25.06~cm (Fig. \ref{fig:schemamanip moteur} (b)) is coupled to a second pulley with a diameter of 70~cm  (Fig. \ref{fig:schemamanip moteur} (c)) thanks to a belt (Fig. \ref{fig:schemamanip moteur} (d)) which allows to multiply the rotational speed by a factor 2.8.
A 3D-printed part, (in green in Fig. \ref{fig:schemamanip moteur}) where ball bearings are inserted, transmit the movement to another pulley (Fig. \ref{fig:schemamanip moteur} (c)), which drives the belt on which the films are formed. Indeed, at the bottom of the setup, placed vertically to this third pulley, there is another identical pulley which is in the liquid bath (Fig. \ref{fig:schema Side Trough} (D)). All of these elements are placed so that the driven belts are vertical and parallel (Fig. \ref{fig:schemamanip} (C)).
The fishing wire (Fig. \ref{fig:schemamanip} (D)) is attached perpendicularly to the vertical belts, thanks to a micrometric screw and a 3D-printed notched ring attached to the belts. The horizontality of the fishing wire has to be carefully controlled. Indeed a curvature of the wire would lead to gradients of drainage along the film during the experiment.

The soap container is a home-made glass tank (Fig. \ref{fig:schemamanip} (G)), with the dimensions of 80~cm long, 12~cm large and 8~cm high. As the length and strain of the 2 vertical belts are fixed, the height of the container is fixed with stands (Fig. \ref{fig:schemamanip} (H)), so that half of the pulley can be immersed in the soapy solution. 

In order to detect the soap films presence, we use a reflection technique. We fix a lighting device (white light torch) on a profile (Fig. \ref{fig:schemamanip} (E)) placed just above the top of the container, at an altitude \textit{d} (see Fig. \ref{fig:schema Side Trough} b)). Three combined light-sensitive resistors (NSL-19M51 purchased from Radiospare) collect the reflected light and assess the presence of a film. During a run, we continuously save the reflected intensity collected by the light-sensitive resistors and compare it to a reference intensity, which corresponds to the room lighting before generating a soap film. As soon as a film is generated and has reached the light-sensitive resistors position, the collected intensity increases and a timer starts until its rupture. We can then obtain the lifetime as well as the maximum length of the soap films (see section \ref{section_FilmsLifetime}). This also allows to automatically start the generation of a new film after the rupture.

A UV-VIS spectrometer (Nanocalc 2000 with a 400~$\mu$m diameter fiber, purchased from Ocean Optics), shown in Fig. \ref{fig:schema Side Trough} (G), is placed on an aluminium profile along the height and we can vary its position (denoted $H$ in Fig. \ref{fig:schema Side Trough} b)). This allows to measure the film thickness, as it is described in section \ref{Section_ThicknessMeasurements}.

When a soap film is entrained, the pulleys rotate in the tank containing the soapy solution: this results in the creation of foam, which changes the conditions of the generation and can even inhibit the formation of soap films. To overcome this problem and to be able to have statistical measurements, we set up an automatic ethanol spray that destroys the foam. To do this, we connected two ethanol sprayers to compressed air using a programmable solenoid valve. We can therefore choose the duration of spraying, as well as its frequency, allowing us to generate films under similar conditions.

\subsection{Interface and Control}

\begin{figure}[htbp]
    \centering
    \includegraphics[width=\columnwidth]{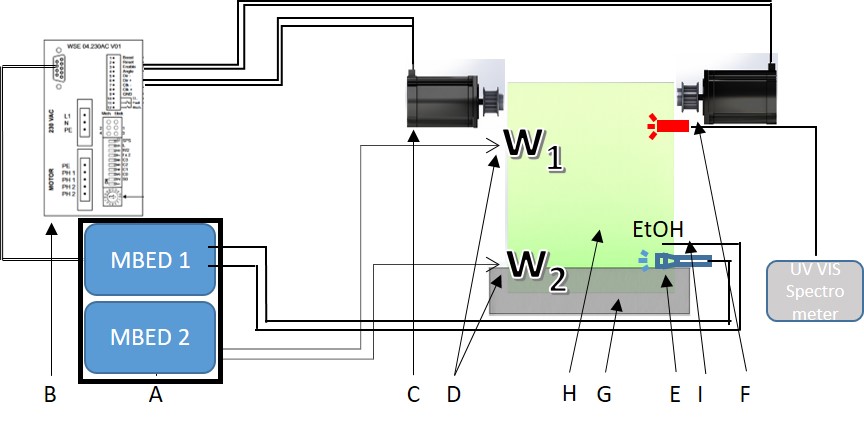}
    \caption{Diagram of the electronic control system. (A) is an electronic box with 2 MBED microcontrollers. The first one controls the variators (B) of the motors (C) and the photoconductive cells (E), the second one controls the humidity captors (D), placed respectively at the top and at the bottom of the film. (F) is the spectrometer, (G) is the tank, (H) represents a soap film. This scheme is not to scale.}
    \label{fig:Elec}
\end{figure}

Most of the components are connected to an electronic box that hosts programmable microcontroller boards (NXP LPC1768 MCU, purchased from Radiospare), which is depicted in Fig. \ref{fig:Elec}. The first board is related to the motors, the reflected light device and the ethanol sprayers, and the second one to the humidity sensors so that there is no conflict with the MCU timers. Using a code editor and a C/C++ compiler, we then create the user interface with LabVIEW. The UV-VIS spectrometer device is a  portable device, controlled by another computer.

At the beginning of an experiment, the user interface allows to redefine the position of the origin (corresponding to the surface of the liquid in the bath) and to generate unique soap films to perform some tests. For statistical analysis, cycles can be conducted, i.e. successive film generation can be initiated automatically. The user can choose the height target, the entrained velocity, the number of cycles and the frequency of use of ethanol sprayers. In each cycle, the lifetime of the soap films, their height as well as the humidity and the temperature in the enclosed box are automatically recorded. 

\begin{figure}[htbp]
    \centering
    \includegraphics[width=\columnwidth]{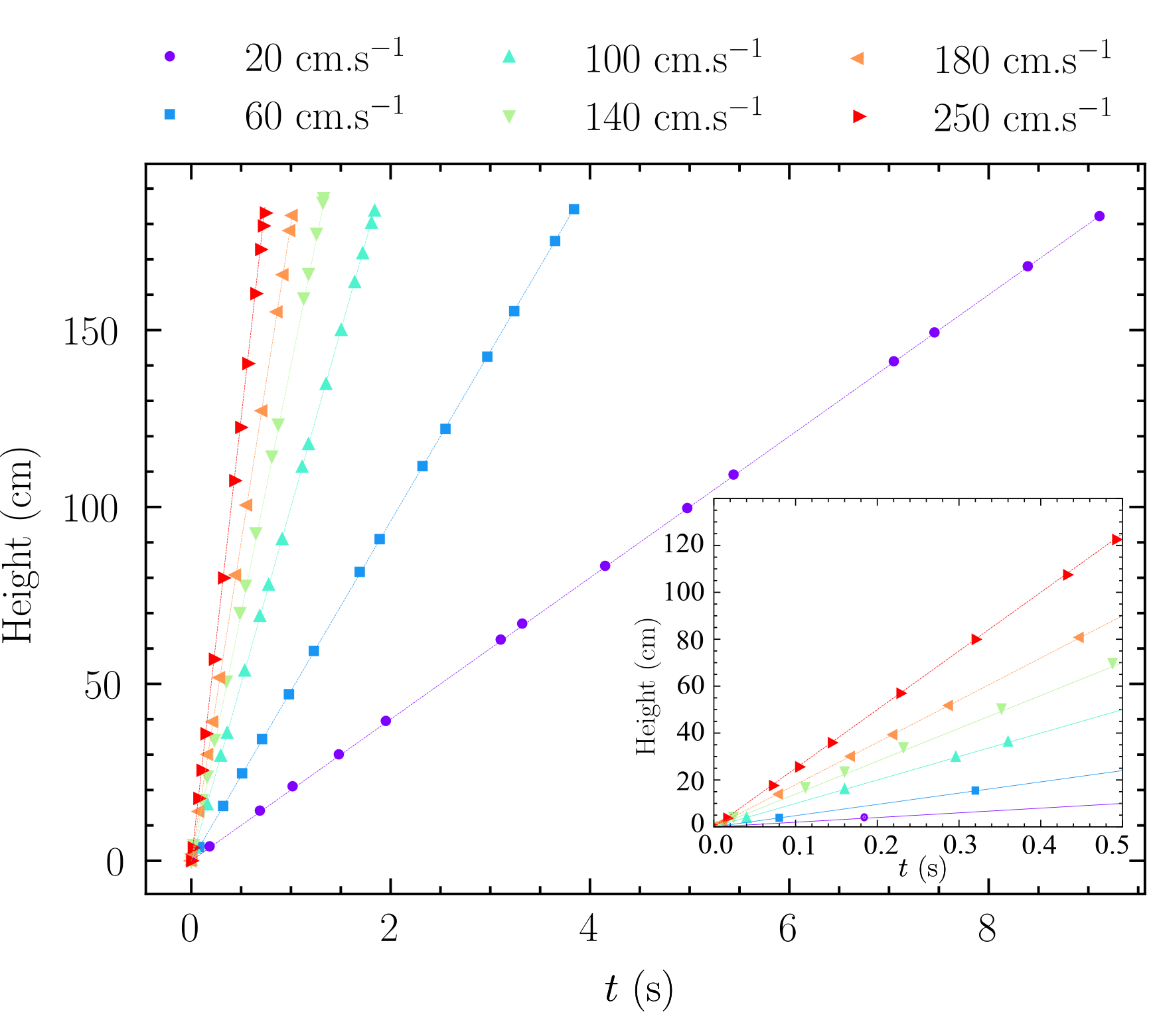}
    \caption{Altitude reached by the fishing line as a function of time. The dots represent measurements made with a high-speed camera with a frame rate equal to 125 for velocities between 20 and 140 cm.s$^{-1}$, and a frame rate equal to 250 for the velocities 180 and 250 cm.s$^{-1}$. The dotted lines correspond to the theoretical height $V \times t$. The inset is a zoom of this graph at short times, for t < 0.5~s. }
    \label{fig:verif_etalonnage-vitesse}
\end{figure}

As rotational motors need an acceleration time, acceleration is made before the fishing line reaches the surface of the soapy solution.  The acceleration phase occurs in the bath, and when the film gets out the entrained velocity is constant. The circular length $l_{c}$ in Fig. \ref{fig:schema Side Trough} represents the added length ran by the fishing line when accelerating. To check that the velocity is actually constant, we have measured for 6 different velocities, the height reached by the fishing line over time using a high-speed camera (Photron Fastcam SA3) at 125 to 250 frames per second. The videos obtained allow to locate the fishing line relatively to the surface of the solution, and to compare this to the expected height, corresponding to the product of the velocity multiplied by time. As can be seen in Fig. \ref{fig:verif_etalonnage-vitesse}, the measured and expected height are in agreement, which allows us to validate the system developed for controlling the entrained velocity.

\section{Controlled environment} \label{Section_Controlled-environnement}

Evaporation has an important impact on the stability of soap films \cite{Champougny2018_evaporation}. It is thus essential to control it in the environment where the films are generated and to be able to measure it over time. 

For this purpose, and in order to protect the films from "draughts", the setup is enclosed with PVC films and PMMA doors. Two room humidifiers, one at the top and one at the bottom of the frame, are used. Humidity and temperature sensors (SHT25 purchased from Radiospare) are present in the chamber to measure the humidity and temperature over time. The humidifier at the bottom (Okoia AH450) allows us to select the target humidity and the humidifier at the top (Bionaire BU1300W-I), which has an adjustable flow rate, allows us to have more homogeneous humidity in the enclosure. This one is set manually at the beginning of an experiment.  

\begin{figure}[htbp]
    \centering
    \includegraphics[width=\columnwidth]{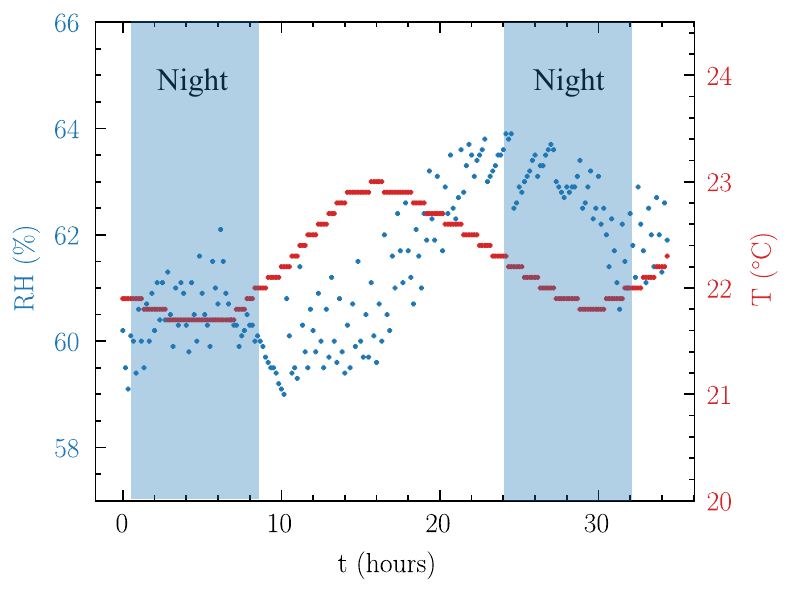}
    \caption{Humidity and temperature recorded over time in the enclosed box for 30 hours, for an experiment performed at RH = 60 $\pm$ 4 $\%$. }
    \label{fig:RH=f(t)}
\end{figure}

If we want to work for example at 60 $\%$ humidity, we can keep it during more than 30 hours (this is longer than the classical experience times) with an absolute accuracy of 4 \% as can be seen in Fig. \ref{fig:RH=f(t)}. The temperature is measured throughout the experiments with an accuracy of 0.1~$^{\circ}$ C, and corresponds to the room temperature which is about 22~$^{\circ}$C.

\section{Measurements} \label{Section_Measurements}

\subsection{Films lifetime} \label{section_FilmsLifetime}

\begin{figure}[htbp]
    \centering
    \includegraphics[width=\columnwidth]{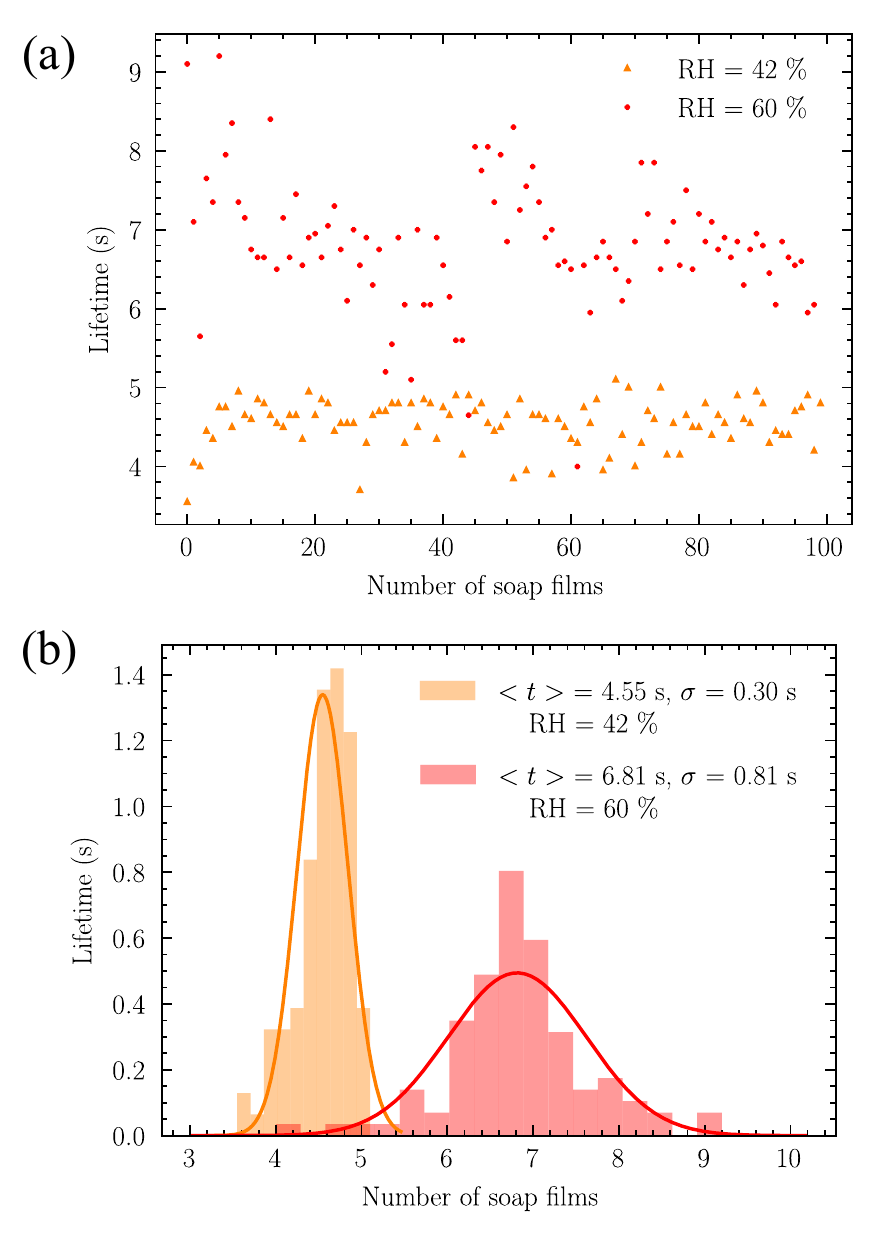}
    \caption{(a) Lifetime measurement of 100 soap films along an experiment performed at a velocity of 20 cm.s$^{-1}$ with controlled humidities of 42 $\%$ and 60 $\%$\ . The soap films break during their generation. (b) Histograms obtained for these two experiments. The lines represent the normal distribution. }
    \label{fig:tdv=f(t)-(a)_histogramme_(b)}
\end{figure}

In this controlled environment, the automatic detection of the films, thanks to the lighting system presented in section \ref{section_Mechanical-parts}, allows us to measure the films lifetime and to ensure their statistical significance. In this paper, all soap films are made with a 4 $\%$ solution of dishwashing liquid (Fairy Original from Procter $\&$ Gamble) dissolved in ultrapure water (resistivity > 18.2~M$\Omega$.cm). Fig. \ref{fig:tdv=f(t)-(a)_histogramme_(b)} (a) shows the lifetime successive measurements of 100 soap films generated continuously at a velocity of 20 cm.s$^{-1}$. The relative humidity is fixed at 42 $\%$ $\pm$ 4 $\%$ (in orange) and 60 $\%$ $\pm$ 4 $\%$ (in red) humidity. The bottom axis indicates the number of the successive films generated during a cycle. For these experiments, soap films break during their generation. The obtained histograms of these measurements are shown in Fig.\ref{fig:tdv=f(t)-(a)_histogramme_(b)} (b). For a given humidity, we measured well-defined distributions. The lines represent the corresponding normal distribution. As expected \cite{Champougny2018_evaporation, miguet_stability_2020}, the average lifetime increases with humidity. For soap films generated at 20 cm.s$^{-1}$, we measure an increase of the lifetime from 4.55~s at 42 $\%$ to 6.81~s at 60 $\%$.

\subsection{Thickness Measurements} \label{Section_ThicknessMeasurements}

\begin{figure}[htbp]
    \centering
    \includegraphics[width=\columnwidth]{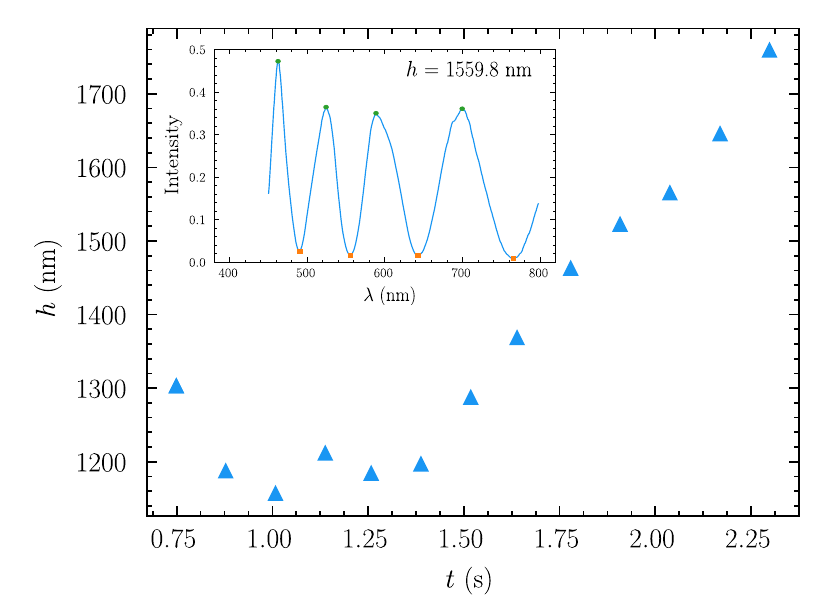}
    \caption{Evolution of a film thickness over time during its generation at a velocity of 60 cm.s$^{-1}$, 13 cm above the soapy solution bath. The inset represents the spectrum of the intensity reflected by a given soap film as a function of the wavelength (blue line: signal; orange squares: minimum detected; green circles: maximum detected). For this spectrum, the film thickness measured is $h =$ 1559 nm.  }
    \label{fig:spectre-(a)_h=f(t)_(b)}
\end{figure}

To study film stability in more detail and to characterize them in the regime of high entrained velocities, it is particularly relevant to be able to measure also their thickness. For that, we used the UV-visible spectrometer described in section \ref{section_Mechanical-parts}. We are thus able to map the thickness of the soap films as a function of the entrained velocity and humidity over time. An example of a spectrum measured for a film thickness $h = 1559$~nm is given in the inset of Fig. \ref{fig:spectre-(a)_h=f(t)_(b)}. The measured spectra are analysed with a Python script. The maxima and minima are detected, which allows to determine the best thickness that fit the data. The spectra are recorded every 100~ms.  

By positioning the spectrometer in the center of the soap film, 13 cm above the soapy solution bath, we recorded its thickness over time during its generation at a velocity of 60 cm.s$^{-1}$. As can be seen in Fig. \ref{fig:spectre-(a)_h=f(t)_(b)}, the soap film thickens over time at this position. This is the first time that this behaviour of films thickening is observed during the generation. This setup will thus allow us to study in more detail the impact of inertia on the stability of soap films.

\subsection{Films visualization}

\begin{figure}[htbp]
    \centering
    \includegraphics[width=\columnwidth]{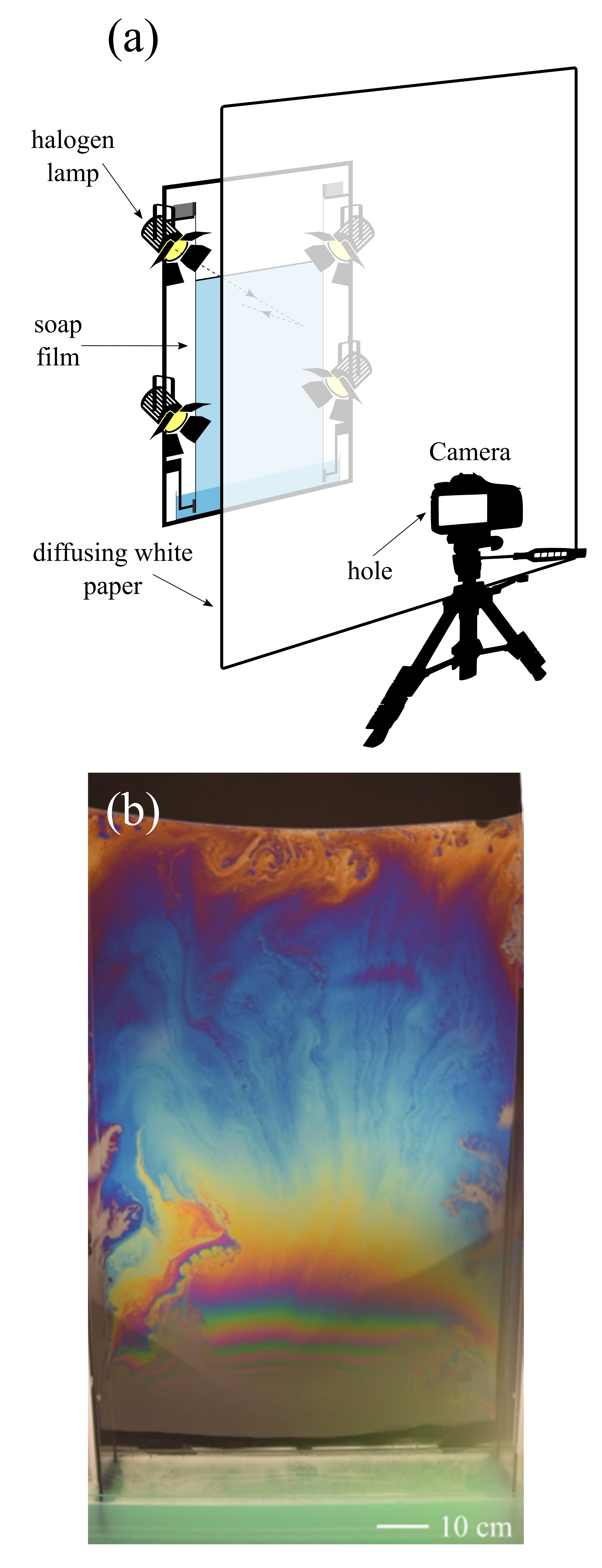}
    \caption{(a) Experimental setup used to visualize soap films. (b) Photograph obtained by using this set-up with a Nikon D7200 camera and a large angle objective lens. The film is generated at a velocity of 50 cm.s$^{-1}$ at a height of 120 cm with a 4 $\%$ Fairy solution.}
    \label{fig:schema_visus_films_geants-(a)_photo-(b)}
\end{figure}

The inhomogeneity of the soap films thickness gives rise to beautiful inhomogeneities of colour on the surface of these films. These colours are due to  light interference between the two air/liquid interfaces of the soap film. Each colour is characteristic of a thickness and the more intense are the colours, the thinner is the film. To visualize this, we installed the setup shown in Fig. \ref{fig:schema_visus_films_geants-(a)_photo-(b)} (a),  which allows us to get photographs from the films. 

To illuminate uniformly the soap films, we use a large white photo studio background paper, which will serve as a reflector. We made a hole in it in order to pass the large angle Nikon lens (AF-P DX 10-20~mm) of a Nikon D7200 camera. We also use 4 halogen lamps, attached to the frame of the setup. The light emitted by these lamps will then be reflected by the large white paper towards the soap films: the colours then appear on their surface, without any parasitic reflection (see Fig. \ref{fig:schema_visus_films_geants-(a)_photo-(b)} (b)).

\section{Conclusion}
In this paper, we presented a new automatized setup for the generation of unfed giant soap films. With it, we can control the height (up to 2 meters) and the entrained velocity (up to 2.5 m/s). This allows us to work in an unexplored regime for the generation of soap films withdrawn by pulling out a frame of a soapy solution, which corresponds to the regime used by the bubbling artists. In parallel, we can also control the ambient humidity and the physico-chemistry of the solutions.
Implemented reflective setup and UV-VIS spectrometer gives information about the lifetime and the thickness of the films. Thanks to automatic repetitions of the experiments, we can access to robust statistical information on the rupture process. These numerous diagnostics allowed by this promising setup will help us to study the stability of giant soap films in a controlled environment.

\section{Acknowledgements}
We acknowledge funding from ESA (MAP Soft Matter Dynamics and contract 4000115113) and CNES (through the GDR MFA). 

\nocite{*}
{Moteurs couplés\bibliographystyle{plain}
\bibliography{biblio}

\end{document}